\documentclass[twocolumn,pra,superscriptaddress]{revtex4}
\usepackage[colorlinks,linkcolor=blue,anchorcolor=blue,citecolor=blue,filecolor=blue,menucolor=blue,runcolor=blue,urlcolor=blue,frenchlinks=blue]{hyperref}
\usepackage{amsmath}
\usepackage{graphicx}
\usepackage{amssymb}
\usepackage{dcolumn}
\usepackage{mathrsfs}
\usepackage{bm}
\usepackage{color}
\begin{document}

\title{Single-modulated-pulse two-qubit gates for Rydberg atoms with noncyclic geometric control}

\author{Zi-Yuan Chen}
\affiliation {Key Laboratory of Atomic and Subatomic Structure and Quantum Control (Ministry of Education), Guangdong Basic Research Center of Excellence for Structure and Fundamental Interactions of Matter, School of Physics, South China Normal University, Guangzhou 510006, China}

\author{Jia-Hao Liang}
\affiliation {Key Laboratory of Atomic and Subatomic Structure and Quantum Control (Ministry of Education), Guangdong Basic Research Center of Excellence for Structure and Fundamental Interactions of Matter, School of Physics, South China Normal University, Guangzhou 510006, China}

\author{Zhao-Xin Fu}
\affiliation {Key Laboratory of Atomic and Subatomic Structure and Quantum Control (Ministry of Education), Guangdong Basic Research Center of Excellence for Structure and Fundamental Interactions of Matter, School of Physics, South China Normal University, Guangzhou 510006, China}

\author{Hong-Zhi Liu}
\affiliation {Key Laboratory of Atomic and Subatomic Structure and Quantum Control (Ministry of Education), Guangdong Basic Research Center of Excellence for Structure and Fundamental Interactions of Matter, School of Physics, South China Normal University, Guangzhou 510006, China}

\author{Ze-Rui He}
\affiliation {Key Laboratory of Atomic and Subatomic Structure and Quantum Control (Ministry of Education), Guangdong Basic Research Center of Excellence for Structure and Fundamental Interactions of Matter, School of Physics, South China Normal University, Guangzhou 510006, China}

\author{Meng Wang}
\affiliation {Key Laboratory of Atomic and Subatomic Structure and Quantum Control (Ministry of Education), Guangdong Basic Research Center of Excellence for Structure and Fundamental Interactions of Matter, School of Physics, South China Normal University, Guangzhou 510006, China}

\author{Zhi-Wei Han}
\affiliation {Key Laboratory of Atomic and Subatomic Structure and Quantum Control (Ministry of Education), Guangdong Basic Research Center of Excellence for Structure and Fundamental Interactions of Matter, School of Physics, South China Normal University, Guangzhou 510006, China}

\author{Jia-Yi Huang}
\affiliation {Key Laboratory of Atomic and Subatomic Structure and Quantum Control (Ministry of Education), Guangdong Basic Research Center of Excellence for Structure and Fundamental Interactions of Matter, School of Physics, South China Normal University, Guangzhou 510006, China}

\author{Qing-Xian Lv}
\email{LQX0801@163.com}
\affiliation {Key Laboratory of Atomic and Subatomic Structure and Quantum Control (Ministry of Education), Guangdong Basic Research Center of Excellence for Structure and Fundamental Interactions of Matter, School of Physics, South China Normal University, Guangzhou 510006, China}

\author{Yan-Xiong Du}
\email{yanxiongdu@m.scnu.edu.cn}
\affiliation {Key Laboratory of Atomic and Subatomic Structure and Quantum Control (Ministry of Education), Guangdong Basic Research Center of Excellence for Structure and Fundamental Interactions of Matter, School of Physics, South China Normal University, Guangzhou 510006, China}

\affiliation {Guangdong Provincial Key Laboratory of Quantum Engineering and Quantum Materials, Guangdong-Hong Kong Joint Laboratory of Quantum Matter, Frontier Research Institute for Physics, South China Normal University, Guangzhou 510006, China}


\begin{abstract}
Arrays of neutral atoms have emerged as promising platforms for quantum computing. Realization of high-fidelity two-qubit gates with robustness is currently a significant important task for large-scale operations. In this paper, we present a convenient approach for implementing a two-qubit controlled-phase gate using Rydberg blockade. We achieve the noncyclic geometric control with a single modulated pulse. As compared with the control scheme by cyclic evolution that determined by dynamical parameters, the robustness of the proposal against systematic errors will be remarkably improved due to the geometric characteristic. Importantly, the noncyclic geometric control reduces the gate time for small rotation angles and will be more insensitive to the decoherence effect. We accelerate the adiabatic control with the aid of shortcuts to adiabaticity to further shorten the operation time. We apply our protocol to the algorithm of quantum Fourier transformation to show the actual acceleration. Therefore, the proposed scheme will provide an analytical waveforms for arbitrary two-qubit gates and may have important use in the experiments of atomic arrays.
\end{abstract}

 \maketitle

\section{Introduction}
Trapped neutral atoms are attractive physical platforms for large scale quantum information systems \cite{Saffman2010,Saffman2016,Henriet2020,Browaeys2020,Scholl2021,Ebadi2021}. Large numbers of atoms can be manipulated in such systems while maintaining excellent quantum coherence. Furthermore, single atom initialization, quantum gates, addressing, and readout have been demonstrated in a variety of optical trapping platforms. Addressable single-qubit gates have been implemented with high fidelity \cite{Xia2015,Wang2016,Wu2019}. Two-qubit gates with neutral atoms can be implemented by driving atoms to highly excited Rydberg states which utilizes the strong and long-range interactions \cite{Jaksch2000}. Many improvements have been made to improve the two-qubit gate fidelity in the past decade, such as employing new Rydberg excitation laser setups to suppress or to eliminate intermediate state scattering, filtering laser phase noise with a high finesse cavity, cooling an atom down to the ground vibrational state in optical tweezers, the optimal control theory, and so on \cite{Maller2015,Jau2016,Levine2018,Leseleuc2018,Liu2021,Madjarov2020}. Recently, high-fidelity parallel entangling gates have been realized on a neutral atomic system \cite{Fu2022,Evered2023}. The next-stage important task will be the realization of addressable two-qubit gates, which would accommodate more robust and faster quantum control.

There are mainly two kinds of two-qubit gates with Rydberg interaction. The first scheme is the Rydberg blockade with the so-called $\pi$-gap-$\pi$ pulses sequences. It has been discussed in \cite{Jaksch2000,Saffman2005,Isenhower2011,Theis2016,Shi2018,Liu2020} and realized in several experiments \cite{Isenhower2010,Wilk2010,Zeng2017,Picken2018,Madjarov}. The fidelity of this scheme will suffer from severe decoherence between the ground state and the Rydberg state which is indicated by the ground-to-Rydberg-state Ramsey oscillations, although the operation period has been shorten to several hundred nanoseconds. The second scheme is using the single-modulated pulses off-resonant modulating driving to drive both the qubit atoms with the same control field \cite{Sun2020,Sun2023,Levine2019}. The population of each two-qubit basis will return to the initial one (cyclic evolution) with phases acquired after the non-adiabatic driving, which can be used to construct the controlled-phase gates. Usually, the dynamics of different two-qubit basis will be asynchrony. Optimization algorithm is adopted to reshape control waveforms to satisfy the cyclic evolution. This features that the operation maybe sensitive to the control paramters which may influence the fidelity of the two-qubit gates in the large-scale arrays. Therefore, realization of robust and high-fidelity two-qubit operations in atomic arrays still need further investigations.

In this paper, we introduce a protocol for the realization of controlled-phase ($\mathrm{C_Z}$) gates in atomic arrays with Rydberg interaction. As in the Rydberg blockade region, different two-qubit basis will experience different dynamical parameters, the geometric control with adiabatic driving would be a natural choice which is robust against random noise and the systematic errors \cite{Berry1984,wilzeck1984,Aharonov1987,Sjoqvist2008,Sjoqvist2016,zhusl2003,duan2001,Sjoqvist2012,Abdumalikov2013,Feng2013,Danilin2018}. However, adiabatic evolution is inherently slow and susceptible to dissipation. To expedite the adiabatic process, shortcut to adiabaticity (STA) methods are considered advantageous. By applying a counter-diabatic Hamiltonian, the errors induced by diabatic effect during fast evolution can be mitigated \cite{Torrontegui2013,Liang2016,felix2018,Yu2018,Yan2019,Vepsalainen2019,Vepsalainen2018,bhhuang,yhchen}. As a step further, the geometric control scheme is generalized to the noncyclic case \cite{Lv2020,Ji2021,Qiu2021,Zhang2021}. The operation time of noncyclic scheme will be much shorter than cyclic one as long as the rotation angle of the two-qubit gates getting smaller. We demonstrate the acceleration results of our proposal in the algorithm of quantum Fourier transformation, a pivotal component of Shor's prime number factoring algorithm. Therefore, our proposal provide a fast and robust way to realize two-qubit gate in atomic arrays with analytical control waveforms.

The structure of this paper is as follows: In section II, The control model of two-qubit gate with Rydberg blockade is introduced. In section III, the proposal of controlled-phase gates with noncyclic geometric control (NCGC) and STA is introduced. In section IV we discuss the robustness of our protocol against random noise and systematic errors and explores its performance under the influence of decoherence. Section V provides a comparative analysis of the time required for the quantum Fourier transform between NCGC and the cyclic case. We summarizes and concludes the paper in section VI.

\section{The control model of two-qubit gate with Rydberg blockade}
In the following we introduce the Hamiltonian of two atoms with Rydberg interaction, where each atom has three levels as labeled by $\{|0\rangle, |1\rangle, |r\rangle\}$, $|0\rangle, |1\rangle$ are the ground states and $|r\rangle$ is the Rydberg state. Ground state $|1\rangle$ of each atom is coupled to Rydberg state with Rabi frequency $\Omega$, detuning $\Delta$ and $\varphi$. The interaction between two Rydberg states of two atoms are given by $V$. Under the single-qubit basis, the total Hamiltonian of the two interacting atoms (as labelled by $1, 2$) is described as
\begin{eqnarray}
&&H=(\frac{\Omega}{2}e^{i\varphi}|1\rangle_1\langle r|\otimes I_2+I_1\otimes|1\rangle_2\langle r|+H.c.)\\
&&+\Delta(|r\rangle_1\langle r|\otimes I_2+I_1\otimes|r\rangle_2\langle r|)+V|r\rangle_1\langle r|\otimes|r\rangle_2\langle r|, \nonumber
\end{eqnarray}
where $I_1=|0\rangle_1\langle 0|+|1\rangle_1\langle 1|+|r\rangle_1\langle r|$, $I_2=|0\rangle_2\langle 0|+|1\rangle_2\langle 1|+|r\rangle_2\langle r|$, $\hbar=1$. By changing to the two-qubit basis, the total Hamiltonian can be divided into three parts
\begin{eqnarray}
&&H=H_1+H_2+H_3,\\\nonumber
&&H_1=(\frac{\Omega}{2}e^{i\varphi}|10\rangle\langle r0|+H.c.)+\Delta|r0\rangle\langle r0|,\\\nonumber
&&H_2=(\frac{\Omega}{2}e^{i\varphi}|01\rangle\langle 0r|+H.c.)+\Delta|0r\rangle\langle 0r|,\\\nonumber
&&H_3=[\frac{\sqrt{2}}{2}\Omega e^{i\varphi}(|11\rangle\langle R|+|R\rangle\langle rr|)+H.c.]\\\nonumber
&&+\Delta|R\rangle\langle R|+(V+2\Delta)|rr\rangle\langle rr|,\nonumber
\end{eqnarray}
where $|R\rangle=(|1r\rangle+|r1\rangle)/\sqrt{2}$. As can be seen that, $|10\rangle(|01\rangle)$ is coupled to $|r0\rangle(|0r\rangle)$ with Rabi frequency $\Omega$ and detuning $\Delta$ while $|00\rangle$ is decoupled from the system. Under the Blockade condition $V\gg\Omega$, $H_3$ can be deduced to
\begin{equation}
H'_3=(\frac{\sqrt{2}}{2}\Omega e^{i\varphi}(|11\rangle\langle R|+H.c.)+\Delta|R\rangle\langle R|,
\end{equation}
with effective Rabi frequency $\sqrt{2}\Omega$. The coupling scheme of $|01\rangle, |10\rangle, |11\rangle$ is shown in Fig. 1(a).

An ingenious method to achieve controlled-phase gate in such system is to drive the system cyclically with phases acquired, i.e., $\varphi_{11}-\varphi_{10}-\varphi_{01}=\pi$, $\varphi_{01}, \varphi_{10}, \varphi_{11}$ are the phases acquired upon states $|01\rangle, |10\rangle, |11\rangle$, respectively. Nevertheless, the dynamics between $|01\rangle(|10\rangle)$ and $|11\rangle$ are different intrinsically (the Rabi frequencies among them are different), and thus optimization algorithm is adopted to design the control waveforms carefully \cite{Sun2020,Sun2023,Levine2019}.

\begin{figure}[ptb]
\begin{center}
\includegraphics[width=8.5cm]{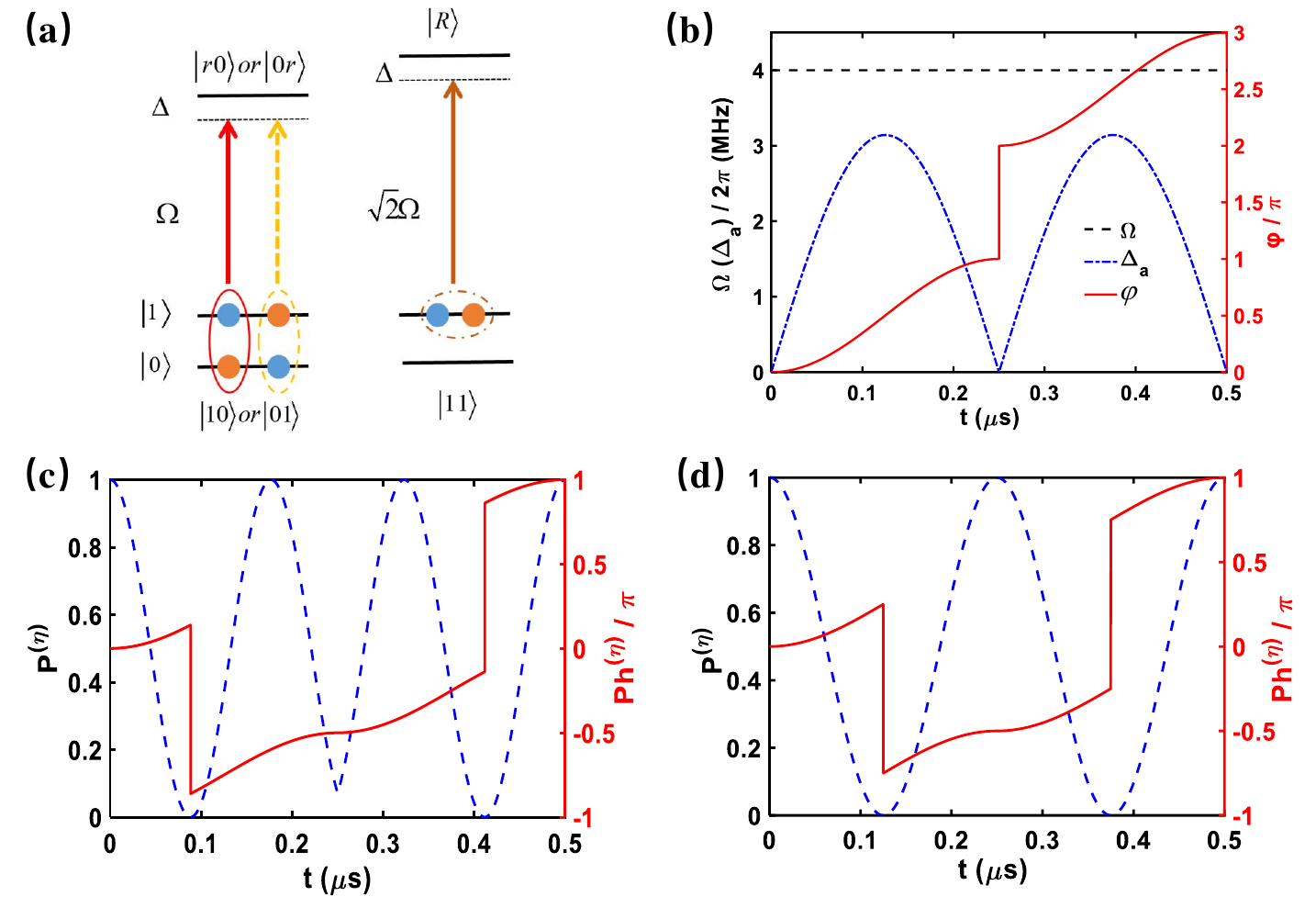}
\caption{\label{fig1}
Scheme of $C_Z$ gate based on noncyclic geometric quantum control. (a) Coupling configuration of two interacting atoms under the two-qubit basis. Under the effect of Rydberg blockade, $|10\rangle (|01\rangle)$ are coupled to $|r0\rangle (|0r\rangle)$ with Rabi frequency $\Omega$ while $|11\rangle$ is coupled to $|R\rangle=(|1r\rangle+|r1\rangle)/\sqrt{2}$ with $\sqrt{2}\Omega$ under the Rydberg blockade condition. All the couplings are shifted by a detuning $\Delta$. $|00\rangle$ is decoupled from the system. (b) Control waveforms of NCGC. Black-dashed line: $\Omega$, red-solid line: $\varphi$. The adiabatic control can be accelerated by adding an auxiliary detuning $\Delta_a$ (blue dashed-dotted line) which induces $\Omega T=4\pi$, $T$ is the evolution period. (c), (d) Dynamics of different basis under the driving. (c) symbols the dynamics of $|11\rangle$ while (d) symbols the ones of $|10\rangle$ or $|01\rangle$. Blue-dashed lines: population $P^{(\eta)}$. Red-solid lines: acquired phases $Ph^{(\eta)}$. As can be seen that all basis return to the initial population with a $\pi$ phase acquired after the driving and thus the $C_Z$ gate is achieved.
}
\end{center}
\end{figure}

\begin{figure*}[ptb]
\begin{center}
\includegraphics[width=15.8cm]{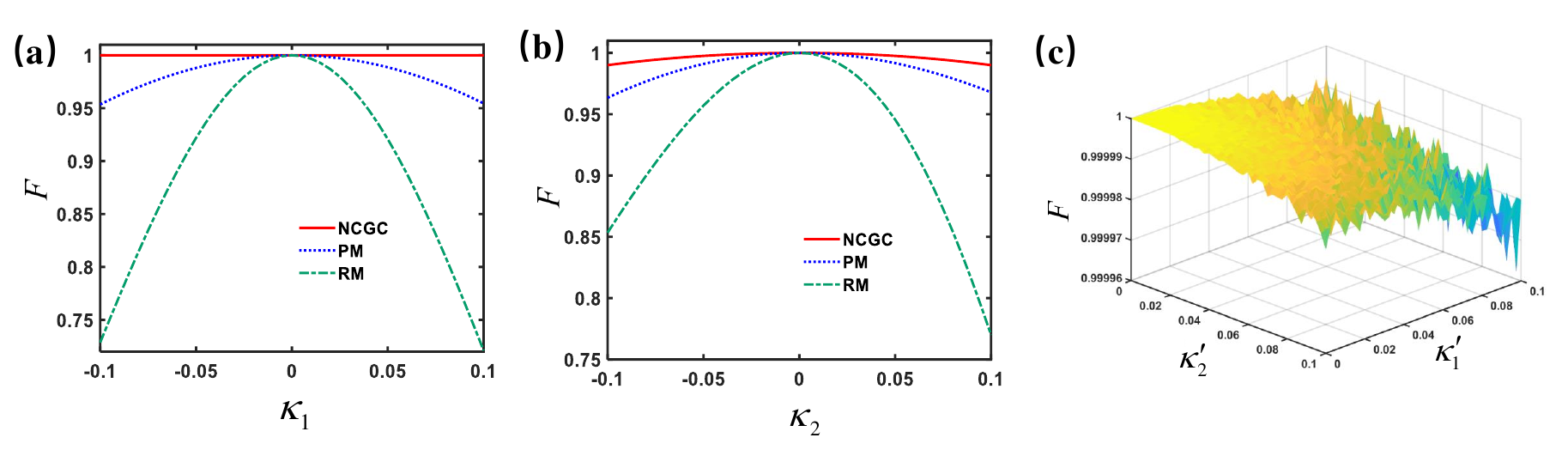}
\caption{\label{fig2}
Robustness of $C_z$ gate realized by NCGC scheme. (a) Fidelity $F$ of $C_z$ gates realized by different protocols against the variation of Rabi frequencies. (a) Fidelity $F$ of $C_z$ gates realized by different protocols against the variation of detuning. Red-solid lines: NCGC, blue-dotted lines: PM, green dashed-dotted lines: RM, where NCGC adopts the waveforms in Eq. (10), RM adopts the ones in Eq. (11) and PM using Eq. (12). (c) Numerical simulation of gate fidelity $F$ against random noise of Rabi frequency and detuning, in which $\Omega_r(t)=\kappa'_1(t)\Omega+\Omega$ and $\Delta_r(t)=\kappa'_2(t)\Omega+\Delta(t)$. $\kappa'_1$, $\kappa'_2$ are time-varying random number and $\kappa'_1 (\kappa'_2)\in[0, 0.1]$. The fidelity of each point was obtained by averaging 50 times.}
\end{center}
\end{figure*}

\section{$\mathrm{C_{Z}}$ gate by noncyclic geometric control with shortcut to adiabaticity}
In the following we introduce how to realize the controlled-phase gate with noncyclic geometric control. Here we introduce the labelling of  $|\mu^{(1)}_{+}\rangle=|r0\rangle$, $|\mu^{(2)}_{+}\rangle=|0r\rangle$, $|\mu^{(3)}_{+}\rangle=|R\rangle$, and $|\mu^{(1)}_{-}\rangle=|10\rangle$, $|\mu^{(2)}_{-}\rangle=|01\rangle$, $|\mu^{(3)}_{-}\rangle=|11\rangle$. Then the eigenstates of Hamiltonian $H_1, H_2, H'_3$ can be written as
\begin{eqnarray}
|\lambda^{(\eta)}_{+}\rangle=\sin(\frac{\theta^{(\eta)}}{2})e^{i\varphi}|\mu^{(\eta)}_{+}\rangle+\cos(\frac{\theta^{(\eta)}}{2})|\mu^{(\eta)}_{-}\rangle,\\ |\lambda^{(\eta)}_{-}\rangle=\cos(\frac{\theta^{(\eta)}}{2})|\mu^{(\eta)}_{+}\rangle-\sin(\frac{\theta^{(\eta)}}{2})e^{-i\varphi}|\mu^{(\eta)}_{-}\rangle, \nonumber
\end{eqnarray}
$\eta=1, 2, 3$, $\theta^{(1)}=\theta^{(2)}=\arctan(\Omega/\Delta)$, $\theta^{(3)}=\arctan(\sqrt{2}\Omega/\Delta)$. The corresponding eigenvalues are derived as $E^{(\eta)}_\pm$=$\pm\hbar\Omega^{(\eta)}$, among which $\Omega^{(1)}=\Omega^{(2)}=\sqrt{\Omega^{2}+\Delta^{2}}, \Omega^{(3)}=\sqrt{2\Omega^{2}+\Delta^{2}}$.

The dynamics of $H_\eta(H'_\eta)$ can be treated as the two-level systems $\{|01\rangle, |r0\rangle\}$, $\{|10\rangle, |0r\rangle\}$, $\{|11\rangle, |R\rangle\}$ governed by control fields with the parameterizations as $\textbf{B}^{\eta}=({\sin\theta^{(\eta)}\sin\varphi, \sin\theta^{(\eta)}\sin\varphi, \cos\theta^{(\eta)}})$. To realize the controlled-phase gate which is independent of $\Omega$, we may adopt $\theta^{(1)}=\theta^{(2)}=\theta^{(3)}=\pi/2$ by setting $\Delta=0$. Namely, we only modulate the phase of the control field. To drive the system evolve adiabatically, we need to satisfy $\dot\varphi\ll \mathrm{Min}[(E^{(\eta)}_{+}-E^{(\eta)}_{-})]/\hbar=2\Omega^{(1)}$, $\mathrm{Min}$ symbols the function that returns the minimal values. Now we assume that the initial state upon each subsystem is given by $|\Psi^{(\eta)}_{0}\rangle=b_1|\lambda^{(\eta)}_+\rangle+b_2|\lambda^{(\eta)}_-\rangle$. The construction of geometric gates can be separated into two stages: firstly, modulating the phase adiabatically from $\varphi^{(0)}$ to $\varphi^{(0)}+\Delta\varphi$, $\varphi^{(0)}$ is the initial phase of the control field. The components of $|\Psi^{(\eta)}_{0}\rangle$ that evolve along $|\lambda_{\pm}^{(\eta)}\rangle$ will pick up a total phase as $\alpha^{(\eta)}_\pm=\delta^{(\eta)}_{\pm}+\beta^{(\eta)}_{\pm}$, where the geometric phases are given by $\delta^{(\eta)}_{\pm}=i\int_{0}^{T}\langle\lambda^{(\eta)}_{\pm}|\frac{d}{dt}|\lambda^{(\eta)}_{\pm}\rangle dt=\pm\Delta\varphi/2$ and the dynamical phases are given by $\beta^{(\eta)}_{\pm}=-\int_{0}^{T} \langle\lambda^{(\eta)}_{\pm}|H_\eta|\lambda^{(\eta)}_{\pm}\rangle dt=\mp\int_{0}^{T}\Omega^{(\eta)} dt$. The evolution operator $U^{(\eta)}_{1}=U(\theta :\pi/2, \varphi:\varphi^{(0)}\rightarrow\varphi^{(0)}+\Delta\varphi)$ under the basis $\{|\lambda^{(\eta)}_{+}\rangle, |\lambda^{(\eta)}_{-}\rangle\}$ is given by \cite{Lv2020}
\begin{equation}
    \label{eq2}
    U^{(\eta)}_{1}=\left(\begin{array}{cc}
         e^{i(\delta_{+}+\beta_{+})}&0  \\
         0& e^{i(\delta_{-}+\beta_{-})}
    \end{array}\right).
\end{equation}
Then we flip the control field by a $\pi$ phase and driving $\textbf{B}^{\eta}$ adiabatically from $\varphi^{(0)}+\pi+\Delta\varphi$ to $\varphi^{(0)}+\pi+2\Delta\varphi$. The evolution operator $U^{(\eta)}_{2}=U(\theta :\pi/2, \varphi:\varphi^{(0)}+\pi+\Delta\varphi\rightarrow\varphi^{(0)}+\pi+2\Delta\varphi)$ is given by
\begin{equation}
    \label{eq3}
    U^{(\eta)}_{2}=\left(\begin{array}{cc}
         e^{i(\delta_{+}+\beta_{-})}&0  \\
         0& e^{i(\delta_{-}+\beta_{+})}
        \end{array}\right).
\end{equation}

Thus, the evolution operator of the union operations is derived as
\begin{equation}
     U^{(\eta)}=U^{(\eta)}_{2}U^{(\eta)}_{1}=\left(\begin{array}{cc}
        e^{i\Delta\varphi} & 0 \\
        0 & e^{-i\Delta\varphi}
    \end{array}\right).
\end{equation}
Here we have used the relationship of $\delta^{(\eta)}_{+}=-\delta^{(\eta)}_{-}$ and $\beta^{(\eta)}_{+}=-\beta^{(\eta)}_{-}$. It can be seen that $U^{(\eta)}=U$ is only depends on the value of $\Delta\varphi$ (has no dependence with $\Omega$) and is thus purely geometric. The proposed scheme is also noncyclic as $\Delta\varphi$ can be set from $0$ to $\pi$. A smaller $\Delta\varphi$ refers to a shorter evolution time $T$ against the same adiabatic condition, and thus the control can be sped up when the rotating angle $\Delta\varphi$ become smaller.

\begin{figure}[ptb]
\begin{center}
\includegraphics[width=6.5cm]{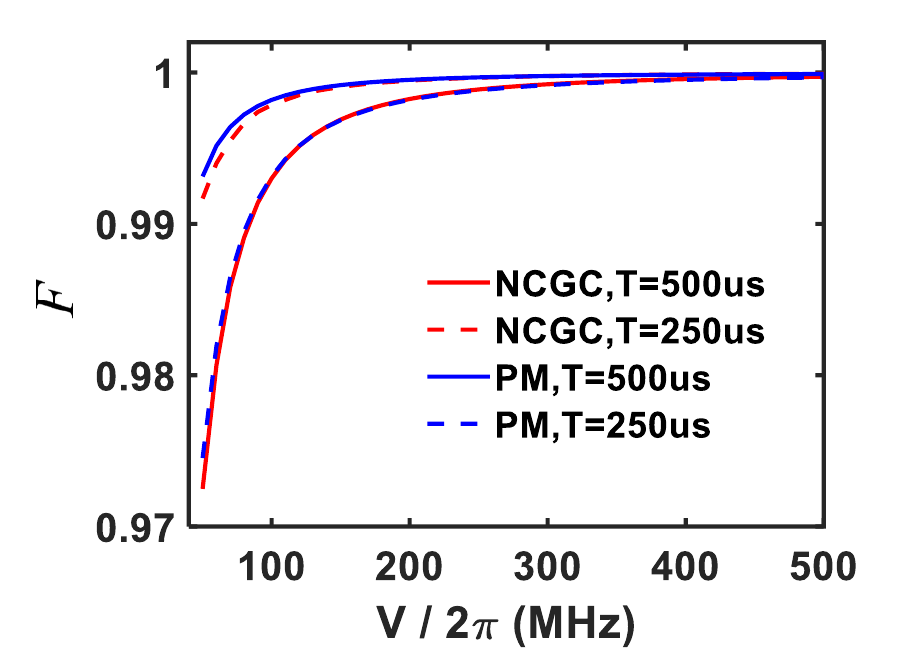}
\caption{\label{fig3}
Fidelity $F$ of $C_z$ gate against Rydberg interacting strength $V$. Red lines: NCGC, blue lines: PM. Solid lines: $T=500 \mu$s, dashed lines: $T=250 \mu$s. The robustness of NCGC scheme against $V$ will be enhanced when the operation time is shorten while the robustness of PM scheme will weaken.}
\end{center}
\end{figure}

Now we turn to the case of two-qubit gate upon the two-qubit basis $\{|00\rangle, |01\rangle, |10\rangle, |11\rangle\}$. According to Eq.(7), states $|01\rangle, |10\rangle, |11\rangle$ will pick up the same phases $\Delta\varphi$ after the noncyclic geometric evolution and the evolution operator will be given by

\begin{equation}
\label{eq7}
    U_{p}=\left(\begin{array}{cccc}
        1 & 0 & 0 & 0\\
        0 & e^{-i\varphi_{01}} & 0 & 0 \\
        0 & 0 & e^{-i\varphi_{10}} & 0 \\
        0 & 0 & 0 & e^{-i\varphi_{11}}
    \end{array}\right)\\,
\end{equation}
where $\varphi_{11}=\varphi_{10}=\varphi_{01}=\Delta\varphi$. By setting $\Delta\varphi=\pi$, we arrive at the $\mathrm{C_{Z}}$ gate that satisfying the condition $\varphi_{11}-\varphi_{10}-\varphi_{01}=\pm\pi$.

To further accelerate the protocol, we adopt the STA scheme through adding auxiliary Hamiltonian $H_{1}^{(\eta)}(t)=i\underset{k=\pm}\sum(|\partial_{t}\lambda^{(\eta)}_{k}\rangle\langle\lambda^{(\eta)}_{k}|-\langle\lambda^{(\eta)}_{k}|\partial_{t}\lambda^{(\eta)}_{k}\rangle|\lambda^{(\eta)}_{k}\rangle\langle\lambda^{(\eta)}_{k}|)$. According to Eq.(4), the expression of $H_{1}^{(\eta)}$ upon the subspace $\{|\lambda^{(\eta)}_{+}\rangle, |\lambda^{(\eta)}_{-}\rangle\}$ will be given by
\begin{equation}
    H_{1}^{(\eta)}(t)=\frac{1}{2}\left(\begin{array}{cc}
       \Delta_a  &  0\\
       0 & -\Delta_a
    \end{array}\right),
\end{equation}
$\Delta_a=\dot\varphi$. As can be seen that all the adiabatic controls and their acceleration in subspace $\{|\lambda^{(\eta)}_{+}\rangle, |\lambda^{(\eta)}_{-}\rangle\}$ can be chosen the same set of control parameters $(\Omega, \Delta, \varphi)$, and thus each two-level system can be manipulated simultaneously.

In Fig. 1(b) we plot the control waveform of controlled-phase gate, where the Rabi frequency is set to be $\Omega_0=2\pi\times4$MHz and detuning $\Delta_0=0$. $V=2\pi\times500$MHz is adopted to confirm the blockade condition. The formulation of phases are given by
\begin{equation}
    \label{eq6}
   \varphi_0=\left\{\begin{array}{lr}
        0.5a\pi[1-\cos(\frac{\pi t}{\tau})],(0<t\leq \tau)&\\
        \pi+0.5a\pi\{3-\cos[\frac{\pi (t-\tau)}{\tau}]\},(\tau<t<T)
   \end{array},\right.
\end{equation}
Here we set a typical evolution period $T=500$ns and $\tau=T/2$ \cite{Evered2023}, $a$ is a constant that determined the rotation angle. To fulfill the condition of $\mathrm{C_{Z}}$ gate, we set $a=1$ ($\Delta\varphi=\pi$). As shown in Fig. 1(c) and 1(d), when the system is prepared upon the state $|\Psi_0^{(\eta)}\rangle=|11\rangle$ ($|01\rangle, |10\rangle$), the population $P^{(\eta)}$ (blue-dashed lines) will return to the initial value after the evolution. The relative phases between the initial states $|\Psi_0^{(\eta)}\rangle$ and the final states $|\Psi_f^{(\eta)}\rangle$ are calculated by $Ph^{(\eta)}=\mathrm{Arg}\langle\Psi_f^{(\eta)}|\Psi_0^{(\eta)}\rangle$, $\mathrm{Arg}$ returns the phase angles of complex numbers. As shown by the red-solid lines in Fig. 1(c) and 1(d), all of the above states will pick up a $\pi$ phase and thus the controlled-phase gate can be achieved. It should be noted that in our calculations $\Omega T=4\pi$ that is not fulfill the adiabatic condition, nevertheless, the system will evolve along the adiabatic pathes after adding the auxiliary Hamiltonian $H_{1}^{(\eta)}$. Therefore, fast and geometric control can be realized in the proposed scheme.

\begin{figure}[ptb]
\begin{center}
\includegraphics[width=5.5cm]{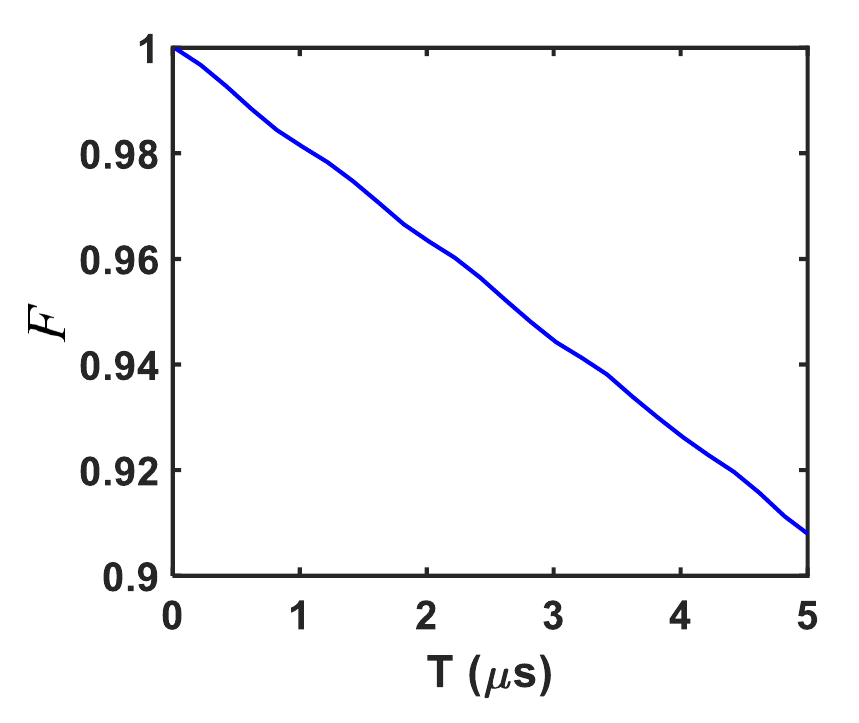}
\caption{\label{fig4}
Numerical simulation of gate fidelity $F$ against the evolution time $T$ under the effect of decay and dephasing. The simulation include the decay from upper energy level $|\lambda^{(\eta)}_+\rangle$ to lower energy level $|\lambda^{(\eta)}_-\rangle$ and upper energy level $|\lambda^{(\eta)}_+\rangle$ to the energy level outside the system $|W\rangle$, as well as a dephasing between the upper energy level $|\lambda^{(\eta)}_+\rangle$ and lower energy level $|\lambda^{(\eta)}_-\rangle$.}
\end{center}
\end{figure}

\section{Performance of $\mathrm{C_{Z}}$ gate of noncyclic geometric scheme}
In the actual experiments, quantum control of the system will be influenced by the offset of the control parameters ($\Omega, \Delta, \varphi$) and the decoherence effect. Here we test the performance of NCGC protocol. In Fig.\ref{fig2} we can study the influence of offset of Rabi frequency and detuning and make a comparison with the other two protocols: single-modulated pulses off-resonant modulating driving \cite{Sun2020} and time optimal gate \cite{Evered2023}. The off-resonant modulating driving scheme can be realized by modulating the Rabi frequency (RM) with a detuning as given by:
\begin{equation}
\begin{aligned}
    &\Omega_1=\sum\limits_{\nu=1}^{4}\beta_\nu[b_{\nu,n}(t/T)+b_{n-\nu,n}(t/T)],\\
    &\Delta_1=2\pi\times3.512 \text{MHz},\\
    &\varphi_1=0,
\end{aligned}
\end{equation}
$\beta_{1}$=1.419 MHz, $\beta_{2}$=0 MHz, $\beta_{3}$=5.076 MHz, $\beta_{4}$=13.425 MHz. $b_{v, n}$ is the $v$th Bernstein basis polynomial of degree $n$ and we set $n=8$. The time optimal scheme is realized by modulating the control phases (PM) as given by:
\begin{equation}
\begin{aligned}
    &\Omega_2= 2\pi\times4.6\, \text{MHz},\\
    &\Delta_2=0,\\
    &\varphi_2=A\cos(\omega t-\varphi_2^{(0)}),
\end{aligned}
\end{equation}
$A= 2\pi\times0.1122$ rad, $\omega=0.1431\Omega_2$, $\varphi^{(0)}_2=-0.7318$ rad. All the coefficients in Eq. (10) and (11) are determined by the optimized algorithm. Due to the finite coherence time of the Rydberg states, we set the evolution time of the three cases (NCGC, PM and RM) to be $T=500$ ns. The Rydberg interacting strength is given by $V=2\pi\times500$ MHz. It can be seen that the control waveforms of NCGC is quite similar to the ones of PM. However, by dividing the operations into two parts, the dynamical phase can be cancelled and the waveforms of arbitrary controlled-phase gates can be obtained analytically. Furthermore, the adiabatic control of NCGC is accelerated by shortcut to adiabaticity which make it less sensitive to the decoherence effect.

In Fig. 2, we compare the robustness of the three schemes (NCGC, PM and RM) against the systematic errors. The fidelity is defined as $F=|tr({U_t}{U_z}^{\dagger})|/16$, $U_z$ is the ideal matrix of $\mathrm{C_Z}$ gate (Eq. (8) with $\varphi_{11}-\varphi_{10}-\varphi_{01}=\pm\pi$) while ${U_t}$ is the evolution operator of the protocols which calculated by $U_t=\mathcal{T}e^{-iH_pt}$, $H_p=H_1+H_2+H'_3$, $\mathcal{T}$ symbols time-ordering operator. $tr$ symbols the trace of the matrix. The deviation of the parameters are introduced by $\Omega'_k=(1+\kappa_1)\Omega_k$, $\Delta'_k=\Delta_k+\kappa_2\Omega_k, k=0, 1, 2$, $\kappa_1 (\kappa_2) \in [-0.1, 0.1]$. The numerical results are shown in Fig. 2(a) and 2(b) where the control parameters are listed in Eq. (10), (11) and (12), red-solid lines: NCGC, blue-dotted lines: PM, green dashed-dotted lines: RM. As can be seen that the noncyclic geometric control does not affected by the deviation of the Rabi frequency since the variation of Rabi frequency only affect the accumulated dynamic phases and the dynamic phase is finally eliminated. The NCGC scheme is also robust against the offset of detuning as shown in Fig. 2(b). Therefore, the intrinsic geometric control is robust against systematic errors as compared with the optimal controls that are only optimized at some special settings.

Here we investigate the robustness NCGC scheme against the random noise. The numerical results are shown in Fig. 2(c) where the random terms are introduced by $\Omega_r(t)=\kappa'_1(t)\Omega+\Omega$ and $\Delta_r(t)=\kappa'_2(t)\Omega+\Delta(t)$. $\kappa'_1$, $\kappa'_2$ are time-varying random number and $\kappa'_1 (\kappa'_2)\in[0, 0.1]$. The parameters settings are the same with Fig. 1. Since the proposed geometric gate is achieved by the accumulation of the noncyclic geometric phases with vanishing dynamical phases,  the fidelity of the protocol keep a fidelity of over 0.9999 even if the amplitude of noise reach 10$\%$ of the Rabi frequency, which shows strong noise robustness.

\begin{figure}[ptb]
\begin{center}
\includegraphics[width=6.5cm]{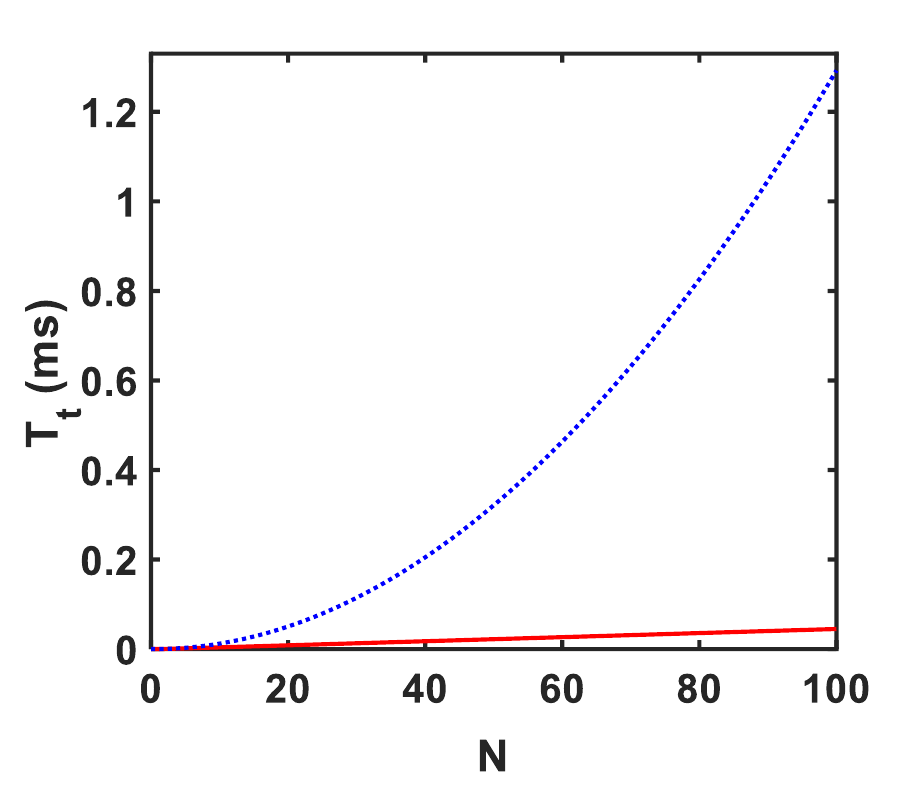}
\caption{\label{fig5}
Numerical simulation of total evolution time of quantum Fourier transformation versus the number of qubits $N$. Blue-dotted line: cyclic scheme, red-solid line: noncyclic scheme. The NCGC scheme rise slower than cyclic scheme as the amount of two-qubit gates with small rotating angles increase polynomially.}
\end{center}
\end{figure}

In Fig.3 we study the influence of Rydberg interacting strength $V$ upon the manipulation. To include the effect of finite interaction strength, we calculate the evolution operator using the fully Hamiltonian, that is, $U_H=\mathcal{T}e^{-iHt}$, $H=H_1+H_2+H_3$.  We make a comparison between NCGC scheme and the PM scheme which adopt the parameters in Eq. (10) and Eq. (12), respectively. When the evolution time $T=500 \mu$s, the fidelity of NCGC (red-solid line) and PM (blue-solid line) will exceed 0.995 when $V>2\pi\times400$ MHz, nevertheless, the NCGC scheme would be more sensitive to the decreasing of strength $V$ as compared with PM. The situation will be different when the evolution time is shorten to $T=250 \mu$s. As can se seen that the robustness of NCGC (red-dashed line) against $V$ will be enhanced while the one of PM (blue-dashed line) drops. One can check that the Rabi frequency of PM scheme should be increase to $\Omega_2=2\pi\times9.2$ MHz in this case to keep $\Omega_2T$ unchange to certificate the cyclic evolution. A larger Rabi frequencies makes the PM scheme more sensitive to $V$. On the other hand, the Rabi frequency of NCGC scheme will be hold as the acceleration is achieved by tilting the detuning $\Delta_a$ in Eq.(9). The decrease of time $T$ would reduce the influence of the imperfect blockade. Therefore, a low ratio of $\Omega_k/V$ at a given evolution time is an essential condition to realize high fidelity two-qubit gate.

In the actual experiment, all the gate protocol will suffer from decoherence. In Fig.\ref{fig4} we simulate the relationship between fidelity of $\mathrm{C_Z}$ gate and the operation time. The simulation include the decay from upper energy level $|\lambda^{(\eta)}_+\rangle$ to lower energy level $|\lambda^{(\eta)}_-\rangle$ and upper energy level $|\lambda^{(\eta)}_+\rangle$ to the energy level outside the system $|W\rangle$, as well as a dephasing between the upper energy level $|\lambda^{(\eta)}_+\rangle$ and lower energy level $|\lambda^{(\eta)}_-\rangle$. We consider a fully Hamiltonian $H=H_1+H_2+H_3$ that include finite Rydberg blockade. We introduce decoherence through master equation which can be written as \cite{Lindblad1976,Johansson2012}
\begin{equation}
\label{eq10}
    \dot\rho=-i[H,\rho]+\underset{n}\sum2L_{n}\rho L_{n}^{\dagger}-L_{n}^{\dagger}L_{n}\rho-\rho L_{n}^{\dagger}L_{n},
\end{equation}
in which $\rho$ is the density matrix of current state, $L_{1}=\sqrt{\gamma_{1}}(|11\rangle\langle R|+|10\rangle\langle r0|+|01\rangle\langle 0r|)$ is decay from upper energy level to lower energy level, $L_{2}=\sqrt{\gamma_{2}}(|W\rangle\langle R|+|W\rangle\langle r0|+|W\rangle\langle 0r|)$ is decay from upper energy level to the energy level outside the system,
$L_{3}=\sqrt{\gamma_{3}}(|R\rangle\langle R|-|11\rangle\langle 11|+|r0\rangle\langle r0|-|10\rangle\langle 10|+|0r\rangle\langle 0r|-|01\rangle\langle 01|)$ is the dephase between upper energy level and lower energy level. Here we adopt a typical dephasing rates as $\Gamma_{1}=1$kHz, $\Gamma_{2}=4$kHz, $\Gamma_{3}=30$kHz and $\gamma_{n}=\Gamma_{n}^2/\sqrt{\Gamma_{1}^2+\Gamma_{2}^2+\Gamma_{3}^2}$ (i.e., Rydberg states with $n=79$) \cite{Liu12021}. The parameters are the same with the ones in Fig. 1. The fidelity is defined as $F=\sqrt{tr(\rho_{ideal}\rho)}$, $\rho_{ideal}$ is the ideal density matrix. As shown in Fig.\ref{fig4}, the fidelity of NCGC will be decrease as gate time increase. To attain a fidelity of over 0.99, the operation time of the protocol should be less than 560ns while a fidelity of 0.999 require a gate time less than 80ns. Given a Rabi frequency $\Omega$ and Rydberg interacting strength $V$, gate time of NCGC can be shortened by tilting detuning $\Delta_a$ in Eq.(9), with the weaken of robustness against the systematic errors.

\section{Acceleration of noncyclic evolution}

In the following we discuss the advantages of using noncyclic evolution. According to the NCGC scheme, the adiabatic condition requires $\dot\varphi\ll2\Omega$, or equivalently, $\Delta\varphi\ll2\Omega T$. Given that $\Omega T=K\Delta\varphi/2$ (Usually, $K\gg10$ to fulfill the adiabatic condition, however, $K\sim1$ after speeding up by the shortcut to adiabaticity), $T=K\Delta\varphi/2/\Omega$. This results that the evolution time $T$ will be shorten by $n$ times when the rotating angle $\Delta\varphi$ is reduced by $n$ times. Therefore, the NCGC scheme offer an analytical method to speed up the control under small angles rotation, without the need of increasing the Rabi frequencies (that has no conflict with the blockade condition $V\gg\Omega$).

As a step further, the NCGC may have potential use in the algorithm of quantum Fourier transformation (the key constitution in Shor's prime number factoring algorithm) of $N$ qubits, where $N$ single-qubit gate (Hadamard gate) and $(L-1)L/2$ two-qubit controlled-phase gate $U_p$ with phase variation of $\Delta\varphi_L=2\pi/2^L$, $L=1, 2, ...,N-1$, are needed. In Fig.5 we calculate the total time that the quantum Fourier transformation cost and make a comparison between the cyclic scheme and the noncyclic one. The controlled-phase gate of noncyclic scheme is realized by the NCGC as described in Fig. 1. Considering the $C_z$ gate and the single-qubit rotation gate $R_{\Delta\varphi}$ form a universal set of quantum computation, the controlled-phase gate $U_p$ of cyclic scheme is accomplished by three gates as \cite{note}
\begin{equation}
U_p=R_{\Delta\varphi}U_zR_{\Delta\varphi}, R_{\Delta\varphi}=\left(\begin{array}{cc}
       \cos\Delta\varphi  &  \sin\Delta\varphi\\
       \sin\Delta\varphi & -\cos\Delta\varphi
    \end{array}\right).
\end{equation}
Here we assume that the operation time of single-qubit gates (Hadamard gates and the rotation gates) and the controlled-phase gate $U_z$ are all set to be $T_N=250$ ns while the operation time of $U_p$ realized by NCGC is decrease by $T_N/(2^L)$ for $\Delta\varphi_L$. As shown in Fig.5, when the number of qubit $N$ increases, the NCGC scheme (red-solid line) shows a great advantage in comparison to cyclic scheme (blue-dashed line) as the amount of two-qubit gates with small rotating angles increase polynomially. Therefore, NCGC provides a much faster way to realize controlled phase gate with small rotation angle, which is of great significance great significance to realize quantum Fourier transformation in large scale of qubit.

\section{Conclusion}
In summary, we have proposed a scheme to realize two-qubit controlled-phase gate with geometric phases in noncyclic evolution. The adiabatic control can be achieved by modulating the phases with constant Rabi frequencies while the acceleration can be tilting the detuning according to shortcut to adiabaticity. The protocol shows a great robustness against the systematic errors, random noise and decoherence. Furthermore, the protocol shows great potentials in the algorithm (i.e., the quantum Fourier transformation) with small rotating angles due to the noncyclic evolution. Therefore, our proposal provide an experimentally feasible and robust way to realize quantum computation in atomic arrays with Rydberg interactions.

\bigskip
\begin{acknowledgments}

This work was supported by the National Natural Science Foundation of China (Grants No. 12074132, No. 12225405, No. 12247123, and No. U20A2074), NSF of Guangdong province (Grant No. GDZX2303006), the China Postdoctoral Science Foundation(Grant No. 2022M721222, No. 2023T160233). Z. Y. Chen and J. H. Liang contribute equally to this work.

\end{acknowledgments}


\begin{thebibliography}{99}
\bibitem{note}It should be noted that altough the optimizating method may give a waveform with shorter time for smaller rotating angle after the optimized algorithm, it seems not the realization of universal quantum computer. Therefore, we use three gates in Eq.(14) to realize the controlled-phase gate $U_p$.
\end{thebibliography}
\end{document}